\documentclass[aps,prl,twocolumn,superscriptaddress,nofootinbib,nobibnotes,floatfix]{revtex4-2}

\usepackage{graphicx}
\usepackage{hyperref}
\usepackage{xcolor}
\usepackage{amsmath}

\begin{document}


\title{Crystallizing Substrates Drag Supported Nanoparticles}



\author{Cheng-Yu Chen}
\affiliation{Department of Materials Science and Engineering, University of Pennsylvania, Philadelphia, PA 19104, USA}
\author{Duncan Burns}
\affiliation{Department of Materials Science and Engineering, Northwestern University, Evanston, IL 60208, USA}
\author{Peter W. Voorhees}
\thanks{Contact author: p-voorhees@northwestern.edu}
\affiliation{Department of Materials Science and Engineering, Northwestern University, Evanston, IL 60208, USA}
\affiliation{Division of Engineering and Applied Sciences, California Institute of Technology, Pasadena CA 91125}
\author{Eric A. Stach}
\thanks{Contact author: stach@engineering.upenn.edu}
\affiliation{Department of Materials Science and Engineering, University of Pennsylvania, Philadelphia, PA 19104, USA}



\date{\today}

\begin{abstract}
When a solid support undergoes crystallization, the advancing amorphous-to-crystalline transformation front separates regions of distinct surface energy, creating a moving interfacial energy boundary. A supported nanoparticle straddling such a boundary experiences an asymmetric particle-substrate interfacial energy environment that constitutes a lateral thermodynamic driving force for migration. Here, using \textit{in situ} transmission electron microscopy to track Pt nanoparticle motion statistically
, paired with time-resolved diffraction and 4D-STEM analysis to characterize support crystallization, we demonstrate that propagating crystallization fronts in amorphous AlO$_x$ thin films actively drag supported Pt nanoparticles over long distances. Temporal correlation between 
the onsets of support crystallization and rapid particle migration, together with 4D-STEM virtual crystallinity maps, establishes that the front drives particle motion. Phase-field simulations confirm that particle-substrate interfacial energy contrast alone sustains particle drag, and identify curvature gradients along the particle surface as the mechanism by which the advancing front redistributes mass and displaces the particle. These results establish a general mechanism by which any propagating surface-energy boundary on a substrate can act as a deterministic driver of supported nanoparticle transport.
\end{abstract}

\maketitle


The motion of solid particles on surfaces driven by surface energy gradients is a fundamental but largely undemonstrated phenomenon. A solid particle whose contact line spans regions of different particle-substrate interfacial energy is not at its lowest energy state, and the system could lower its total energy by displacing the particle toward the region of higher particle-substrate adhesion energy. In contrast, the analogous problem for liquid droplets is well understood: Brochard showed theoretically that a spatial gradient in the spreading coefficient causes droplets to migrate toward regions of higher wettability,\cite{brochard1989motions} and Chaudhury and Whitesides demonstrated this experimentally, driving water droplets uphill against gravity on a chemically modified surface.\cite{chaudhury1992make} In these systems, the driving force acts through an imbalance in capillary forces at the contact line, and the response rate is set by viscous dissipation inside the droplet. However, demonstrating the same effect for solid particles is challenging: unlike a liquid droplet, a solid particle responds to a contact-line imbalance only through thermally activated surface diffusion,\cite{mullins1957theory} which is orders of magnitude slower than viscous relaxation and operative only for sufficiently small particles at sufficiently high temperatures. Furthermore, as particles grow through sintering, the increasing material redistribution required to achieve the same center-of-mass displacement makes sustained gradient-driven migration progressively more difficult.\cite{hansen2013sintering} As a result, gradient-driven center-of-mass motion of solid particles on substrates has remained largely unexplored.

Techniques used to describe liquid migration can be expanded to solid systems. Raphael {\it et al.} developed a theory for center of mass velocity ($v_{cm}$) of a cylindrical wedge segment driven by surface energy difference, following experiments conducted by Brochard {\it et al.} on static substrates \cite{raphael1989capillary, brochard1989motions}. The liquid-state approach utilizes a combination of mass conservation and entropy production constraints by equating the entropy production rate around equilibrating contact lines to the viscous dissipation. In the solid-state, particles are assumed near their equilibrium contact angles ($\theta_{eq}$). When atop a heterogenous substrate, entropy can be produced by particles moving toward the substrate with higher adhesion energy \cite{chen2025interfacial}. With constant particle surface area, a macroscopic configurational force can be expressed by, $F_c = \gamma^{v}_{NP}(\cos(\theta_{Y, +\hat{x}}) - \cos(\theta_{Y, -\hat{x}}))\partial A^{NP}_{+\hat{x}} / \partial x$, in terms of particle-vapor surface energy ($\gamma^{v}_{NP}$), equilibrium contact angles ($\theta_{Y, \pm\hat{x}}$) for each substrate, and the increase in particle-($+\hat{x}$) substrate area as the particle moves toward the +$\hat{x}$ direction. Considering a cylindrical wedge geometry, $\partial A^{NP}_{+\hat{x}} / \partial x = l$, the length of the wedge, thereby allowing simplification by instead using entropy production per unit length. For a wedge moving with center of mass velocity, $v_{cm}$, the entropy production rate per unit length can be expressed via $F_{\text{config}}/l \cdot v_{\text{cm}}$. This macroscopic rate can then be equated to the microscopic entropy gain from adatom diffusion, with adatom current $j$, and mobility, $M$, as $\int dV \left[j^2 / M\right]$, where integration over volume, $V$, accounts for all currents of the particle species along a slice of the wedge. This expression accounts for bulk-, surface-, and phase-change-driven flows, but is subject to particle mass conservation. 
For solid particles far from their transition point, surface diffusion dominates the integration. Mass flow then occurs through a shell of surface thickness, $\delta$, associated with the decay length of the mobility and long-range disjoining potentials emanating from the solid surface. The surface diffusion can be captured via, $j = -(D_s\nu_s / k_B T) \nabla_s \mu$, with diffusion coefficient, $D_s = M k_B T$, surface density, $\nu_s$, and surface chemical potential $\mu$. When curvature ($\kappa$) variations dominate, $\nabla\mu = \Omega \nabla(\gamma \kappa)$, and entropy balance can be reduced to,
\begin{equation}\label{eqn_veli}
    \begin{split}
        &\gamma^{v}_{NP}(\cos(\theta_{Y, +\hat{x}}) - \cos(\theta_{Y, -\hat{x}})) v_{\text{cm}} \\
        &= \left( \frac{D_s \nu_s^2 \Omega^2 \gamma \delta}{k_B T}\right)\int dS \left[|\nabla_s\kappa|^2 \right],
    \end{split}
\end{equation}
with atomic volume, $\Omega$. We identify an effective migration velocity scale $v^* = \Gamma\left(\frac{D_s \nu_s^2 \Omega^2 \gamma \delta}{k_B T R^2}\right)$ for a particle of radius $R$. For a $5$ nm platinum particle at $800^\circ C$, $v^* = \Gamma\cdot\mathcal{O}(10^{5})$nm$ / s$, computed using coefficients specific to Pt \cite{blakely1962, lee2018}, which is multiplied by a dimensionless curvature gradient factor, $\Gamma$, setting the final velocity. $v^*$ scales as $1/R^2$, implying that capillarity-driven motion is possible for nano-scale particles. While $v^*$ provides an estimate for the velocity, there can be complex adatom dynamics at the contact lines and the particle-substrate contact that have not been accounted for. In particular, $\nabla\mu$ can include temperature, stress, and composition gradients that may lead to particle motion.
We demonstrated in our previous work that sufficiently large interfacial energy contrasts on static-heterogeneous substrates can drive supported Pt nanoparticles along deterministic migration trajectories, \cite{chen2025interfacial} establishing that the effect is experimentally accessible. However, a static gradient produces only a finite particle displacement toward a new equilibrium position, and whether a propagating surface-energy boundary can sustain such motion continuously remains an open question.

Crystallization of a nanoparticle support can provide such a propagating boundary. When a support undergoes an amorphous-to-crystalline transformation, the advancing front continuously separates regions of distinct surface energy, and any particle straddling the front sits in an asymmetric interfacial energy environment that favors displacement toward the region of higher adhesion energy. Here, using \textit{in situ} TEM paired with time-resolved diffraction and 4D-STEM analysis, we demonstrate that propagating crystallization fronts in amorphous AlO$_x$ thin-film supports actively drag supported Pt nanoparticles over long distances. Amorphous alumina was selected as the support because its crystallization sequence is well defined \cite{nayar2014structural,dragoo1967transitions} and its transformation temperature can be tuned through film thickness,\cite{mavrivc2019advanced,tavakoli2013amorphous,edlmayr2010thermal} making it a controllable model system for studying front-particle coupling. Small Pt nanoparticles were selected because of their large surface curvature \cite{Jeyaraj2019}. The thin-film geometry enables the positions of many particles to be tracked, while time-resolved diffraction captures the onset and progression of crystallization of the support.


\begin{figure*}
    \centering
    \includegraphics[width=\textwidth]{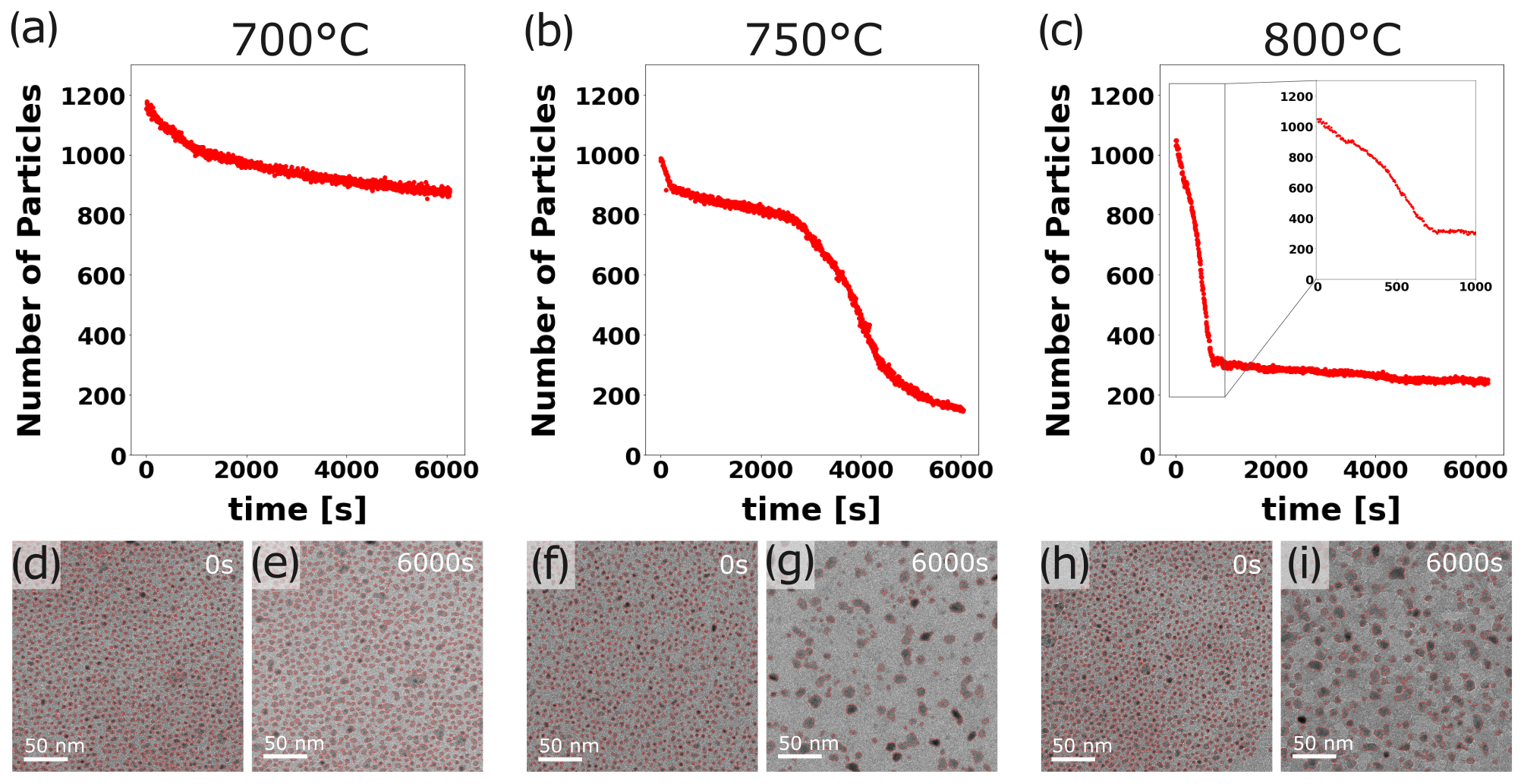}
    \caption{Distinct temperature-dependent regimes in the evolution of Pt particles on AlO$_x$ supports during \textit{in situ} TEM heating. (a--c) Particle number as a function of time during heating at 700, 750, and 800 $^\circ$C, respectively. (d--i) Representative TEM images recorded at 0 and 6000 s at each temperature: (d,e) 700 $^\circ$C, (f,g) 750 $^\circ$C, and (h,i) 800 $^\circ$C. Red outlines mark particle boundaries.}
    \label{fig1}
\end{figure*}


To investigate the dynamic evolution of the cyrstallization of the AlO$_x$ support   we first deposit amorphous AlO$_x$ thin films onto heating chips~\cite{hummingbird} using an electron-beam physical vapor deposition (PVD) system (see Supplemental Information for further details). To verify that AlO$_x$ crystallization is a thermal process rather than a consequence of electron-beam irradiation or metal--support interactions, an bare AlO$_x$ thin film was heated to $800^\circ$C for 90 min in the TEM. Selected-area electron diffraction (SAED) patterns were collected every 5 min during heating from a region under continuous electron irradiation, and once after heating from an adjacent region that had received no prior beam exposure. Radially integrated intensity profiles of the SAED patterns (Fig.~S1(a)) show reflections associated with crystalline $\gamma$- and $\delta$-Al$_2$O$_3$ in both regions at $800^\circ$C. This result is consistent with previous reports showing that electron-beam-evaporation-grown amorphous alumina films crystallize into $\gamma$- and $\delta$-Al$_2$O$_3$ at this temperature.\cite{nayar2014structural} TEM imaging together with 4D-STEM data (Fig.~S1(b,c)) further confirms the formation of dendritic nuclei from crystalline alumina phases. Because crystallization is observed in both regions at a beam dose rate far below values reported to induce AlO$_x$ crystallization,\cite{murray2012electron} these results indicate that AlO$_x$ crystallization is thermally driven and occurs independent of both electron-beam-induced effects and Pt particles. Heterogeneous nucleation at surface defects inherent to the as-deposited film may, however, influence the local onset and rate of crystallization.


To examine how the transformation of the support influences the evolution of  nanoparticles, Pt particles were deposited onto freshly prepared 10~nm AlO$_x$-coated heating chips and monitored during \textit{in situ} TEM heating at 700, 750, and 800$^\circ$C. A U-Net-based image segmentation workflow was applied to the time-resolved TEM images to track the number of particles as a function of time \cite{Horwath2020,Vyas2022}. At all three temperatures, an initial rapid drop in particle number reflects coalescence among closely spaced particles. At 700$^\circ$C, after the initial stage, most particles remain near their original locations, and the number of particles only gradually decreases 
over 100~min (Fig.~\ref{fig1}(a,d,e) and Supplemental Movie 1). At 750 and 800$^\circ$C, however, a second and more dramatic drop occurs after a temperature-dependent delay
. At 750$^\circ$C this transition begins at $\sim$3000~s, whereas at 800$^\circ$C it occurs much earlier ($\sim$250~s) and more abruptly
. In both cases, rapid migration eventually ceases and the particle number stabilizes, consistent with particles becoming immobilized once the support has largely crystallized (Fig.~\ref{fig1}(b--i) and Supplemental Movies 2--3).

Post-heating SAED analysis confirms the link to  crystallization of the support: no crystalline alumina reflections are detected after heating at 700$^\circ$C, while reflections consistent with $\gamma$-, $\theta$-, and $\alpha$-Al$_2$O$_3$ are present after heating at 750 and 800$^\circ$C (Fig.~S2). The appearance of additional crystalline polymorphs beyond the predominantly $\gamma$-Al$_2$O$_3$ observed for the empty film may reflect heterogeneous nucleation at Pt/support interfaces. The coincidence between the onset of rapid particle migration and the crystallization of the  support  at 750 and 800$^\circ$C, and its absence at 700$^\circ$C where the support remains amorphous, indicates that crystallization actively drives particle migration through contact-line 
asymmetry imposed by the advancing crystallization front \cite{chen2025interfacial}. Based on these observations, we identify four distinct regimes of particle evolution: 

\begin{enumerate}
    \item Initial coalescence among closely spaced particles, present at all temperatures
    \item Limited migration on the amorphous support, which dominates at 700$^\circ$C and precedes the onset of crystallization at higher temperatures
    \item Rapid migration driven by the nucleation and growth of crystalline Al$_2$O$_3$, observed only at 750 and 800$^\circ$C
    \item Particle locking as growing crystallites impinge and cover most of the support surface
\end{enumerate}

Of the four regimes identified above, the third provides the clearest opportunity to test our mechanistic hypothesis, because it captures the interval during which rapid particle migration occurs while the support is actively crystallizing. We therefore focus on this regime to examine whether the onset of particle migration coincides  with the onset of support crystallization by collecting SAED patterns and corresponding TEM images during \textit{in situ} heating.

\begin{figure*}
    \centering
    \includegraphics{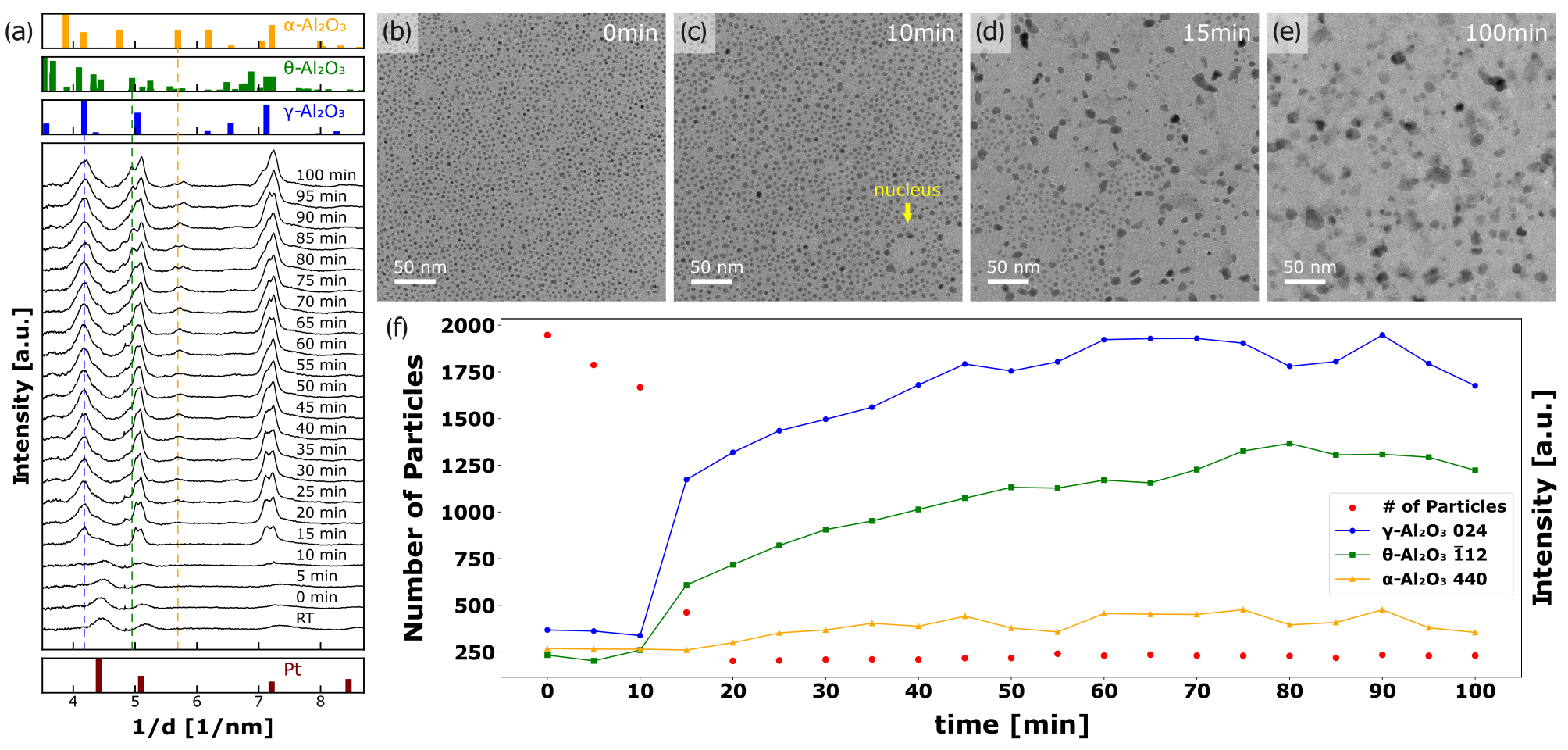}
    \caption{Temporal correlation between support crystallization and particle migration during \textit{in situ} TEM heating at $800^\circ$C. (a) Time-resolved radially integrated intensity profiles of SAED patterns collected during heating of Pt particles on a 20 nm AlO$_x$ thin film at $800^\circ$C. Reference reflection positions for $\alpha$-, $\theta$-, and $\gamma$-Al$_2$O$_3$ and Pt are indicated for comparison. Colored dashed lines indicate the representative diffraction reflections assigned to $\gamma$-Al$_2$O$_3$ 024, $\theta$-Al$_2$O$_3$ $\bar{1}$12, and $\alpha$-Al$_2$O$_3$ 440, whose integrated intensities are plotted in (f). (b--e) Representative TEM images acquired after 0, 10, 15, and 100 min of heating, respectively. (f) Time-dependent changes in particle number and selected diffraction intensities associated with crystalline alumina reflections.}
\label{fig2}
\end{figure*}


To make the onset of particle migration easier to identify, we sought to reduce the number of crystallization nuclei so that particles could migrate over longer distances before becoming trapped between neighboring alumina domains. For this reason, a thicker 20~nm AlO$_x$ film was used, as greater film thickness can close pinhole openings during PVD deposition~\cite{panjan2020review}, reducing the density of potential nucleation sites. Atomic force microscopy (AFM) maps of the as-deposited 10 and 20~nm AlO$_x$ films confirm that the 20~nm film contains substantially fewer pinholes than the 10~nm film (Fig.~S3).


Pt particles were then deposited onto the 20~nm AlO$_x$ film, and the sample was heated to 800$^\circ$C while SAED patterns and corresponding TEM images were acquired every 5~min over a 100~min heating period. The time-resolved SAED data show that crystalline alumina reflections begin to emerge after approximately 10~min of heating and become clearly pronounced after approximately 15~min (Fig.~\ref{fig2}(a)). This timing closely coincides with the onset of pronounced Pt particle migration in the corresponding TEM images. At the start of the acquisition period (0~min), the particles remain well dispersed on the support (Fig.~\ref{fig2}(b)). After 10~min of heating, the overall particle distribution is still largely preserved; however, a localized region can be identified from which nearby particles appear to migrate outward from a common center (Fig.~\ref{fig2}(c), yellow arrow). Although a crystalline domain cannot be identified directly from the TEM image alone, the appearance of this localized outward particle motion near the time when crystalline alumina reflections first emerge in SAED suggests that this region corresponds to an early crystallization nucleus. By 15~min, particle migration and coalescence become much more pronounced (Fig.~\ref{fig2}(d)), and after 100~min the surface is populated by substantially larger coalesced particles separated by broad particle-free regions (Fig.~\ref{fig2}(e)).

\begin{figure*}
    \centering
    \includegraphics{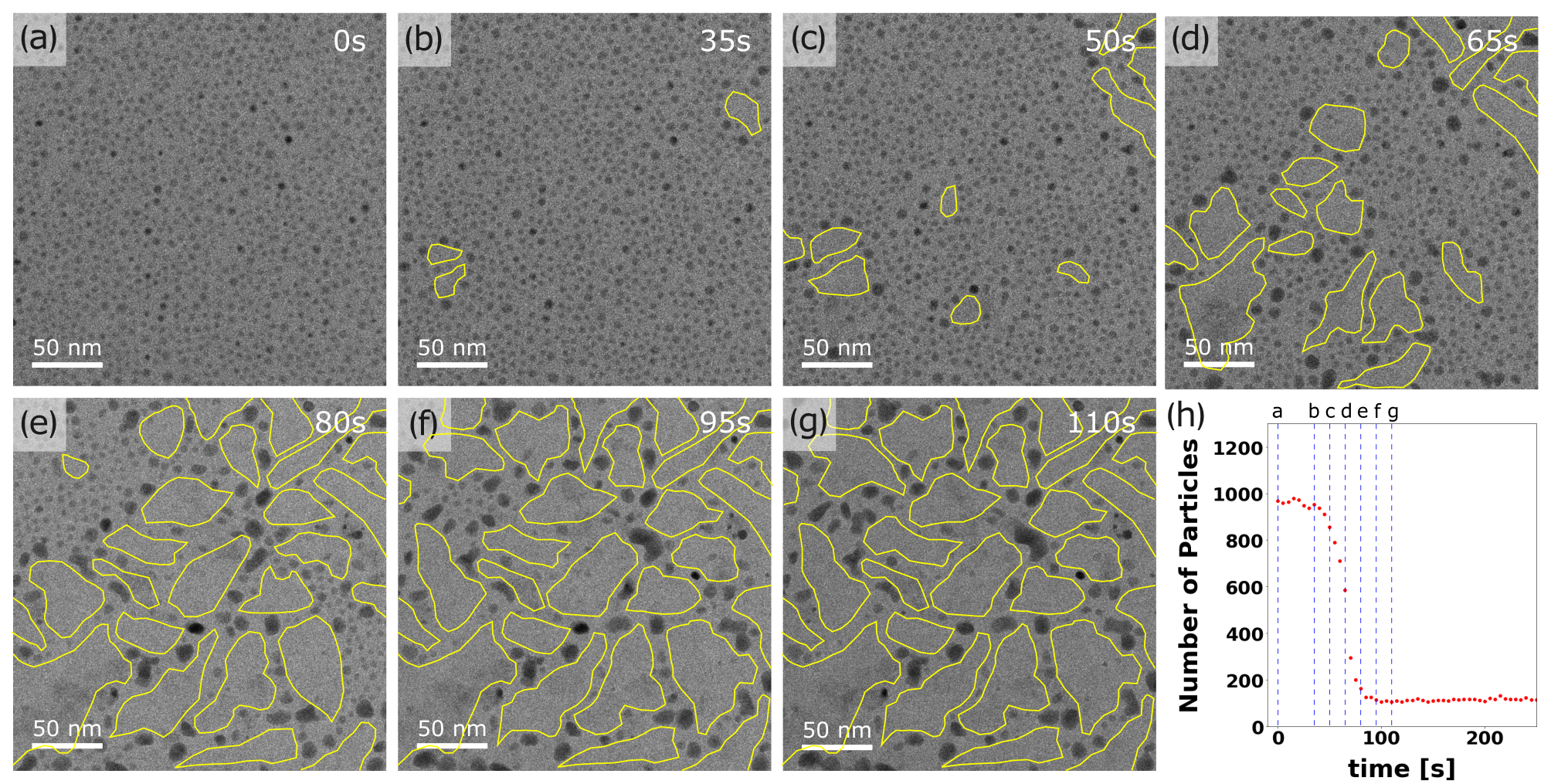}
    \caption{Dynamic evidence linking Pt particle migration to the lateral propagation of support crystallization during \textit{in situ} TEM heating at $800^\circ$C. (a--g) Time series from a separate high-temporal-resolution \textit{in situ} TEM heating experiment on a 20 nm AlO$_x$ thin film. Yellow contours outline particle-free regions. (h) Particle number as a function of time. Blue dashed lines indicate the time points corresponding to panels (a--g).}
\label{fig3}
\end{figure*}

The slight offset between the image-based onset of particle migration (10~min) and the more clearly developed crystalline reflections in SAED (15~min) may arise for two reasons. First, the earliest crystal nuclei may be too small to generate diffraction intensity sufficiently strong to be clearly resolved in the SAED profiles. Second, each  the SAED pattern was acquired slightly before the corresponding TEM image. 

Quantitative analysis further shows that the apparent increase in diffraction intensities associated with crystalline alumina reflections occurs concurrently with the sharp decrease in particle number (Fig.~\ref{fig2}(f)). Three representative diffraction reflections associated with $\gamma$-, $\theta$-, and $\alpha$-Al$_2$O$_3$ were selected to illustrate the intensity evolution over time more clearly. After the pronounced increase between 10 and 15~min, the intensities of these reflections continue to rise as the crystalline domains grow. We also note that diffraction reflections associated with the additional crystalline polymorphs $\theta$- and $\alpha$-Al$_2$O$_3$ are more pronounced for the 20~nm film (Fig.~\ref{fig2}(a)) than for the 10~nm film (Fig.~S2, 800$^\circ$C). Several previous reports have shown that as the specific surface area decreases, $\alpha$-Al$_2$O$_3$ becomes more energetically stable than $\gamma$-Al$_2$O$_3$ and amorphous AlO$_x$~\cite{mavrivc2019advanced,mchale1997surface,drazin2017reducing,tavakoli2013amorphous}, which may explain the stronger $\alpha$-Al$_2$O$_3$ reflections observed for the 20~nm film.

Taken together, these observations indicate that the onset of Pt particle migration is temporally aligned with the onset of crystallization of the support, consistent with particle migration being driven by the propagating crystallization front.


To further probe the dynamic interaction between support crystallization and particle motion, we perform a separate \textit{in situ} TEM heating experiment at 800$^\circ$C with higher temporal resolution captured every 5 s (Fig.~\ref{fig3}). The contours shown in Fig.~\ref{fig3}(b--g) outline regions from which Pt particles have migrated away, leaving particle-free areas behind. Although the crystalline domains themselves cannot be directly identified from these TEM images alone, the progressive expansion of these particle-free regions provides indirect dynamic evidence for a propagating crystallization front that drags nearby particles outward. As the outlined regions grow, adjacent Pt particles undergo pronounced migration and coalescence, accompanied by a rapid decrease in particle number (Fig.~\ref{fig3}(h)). This behavior is consistent with the hypothesis that Pt particle motion is driven by the advancing crystallization front.

\begin{figure*}
    \centering
    \includegraphics[width=\textwidth]{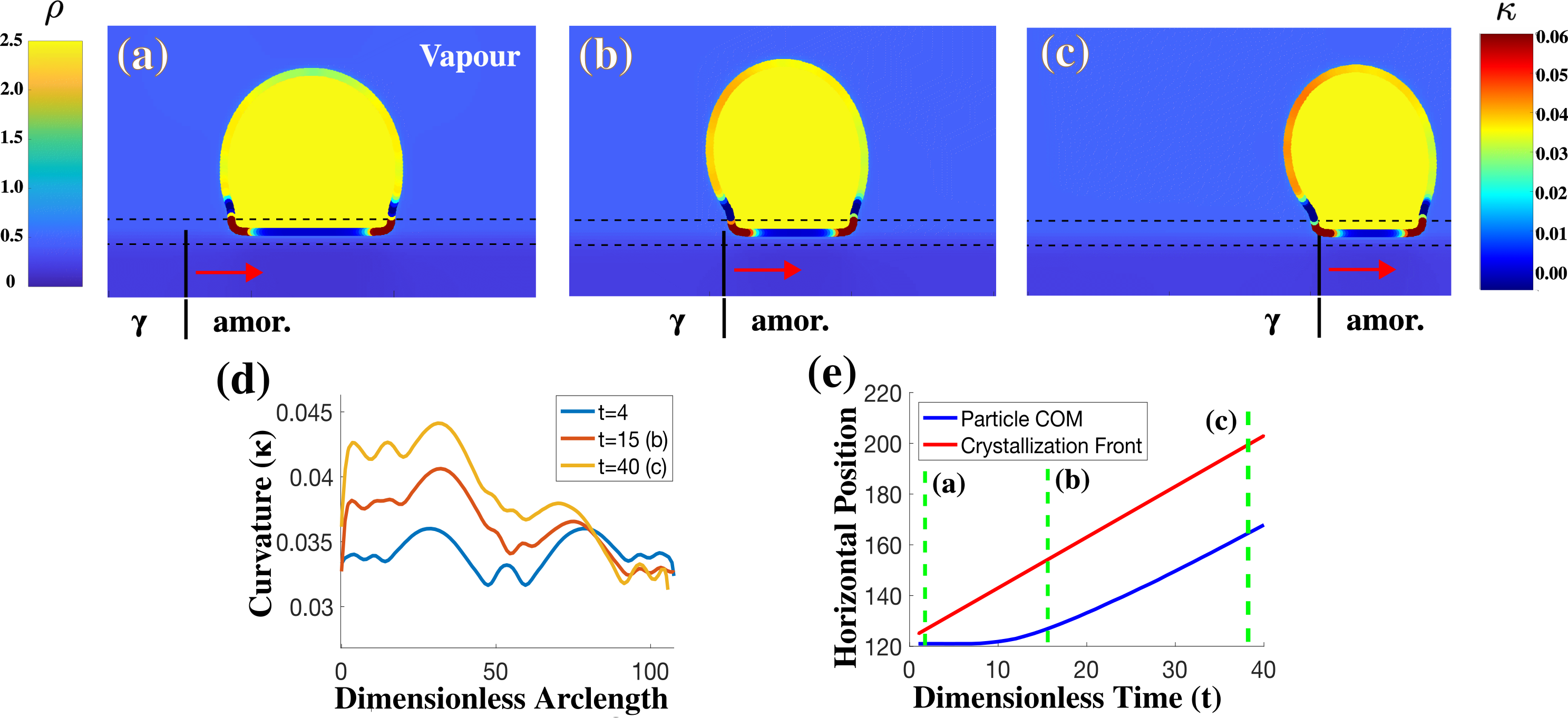}
    \caption{Simulation of crystallization front induced particle motion. (a-c) Pt density map snapshots. Particle curvature ($\kappa$) is plotted along the platinum solid-vapor interface ($\phi=0.5$). (d) Particle curvature as a function of arclength from left to right above the contact point core region. (e) Center of mass position of particle (blue) and crystallization front position (red) as a function of simulation time, $t = 10^5 dt$.}
\label{fig4}
\end{figure*}

As evolution proceeds, neighboring particle-free regions eventually meet, after which their boundaries show little further change. At the same time, Pt particle migration becomes strongly suppressed, and the number of particles reaches a plateau with only minimal variation thereafter (Fig.~\ref{fig3}(h)). The alignment between the cessation of expansion of the particle-free regions and the stabilization of particle number is again consistent with our hypothesis: once the crystallization fronts impinge on one another and can no longer propagate freely, the driving force for long-range particle migration is strongly reduced.


In Fig.~\ref{fig3}, some of the contoured particle-free regions exhibit darker contrast in the bright-field TEM images. This observation is qualitatively consistent with diffraction contrast from crystalline material, since crystallized regions can scatter electrons away from the bright-field beam more strongly than the surrounding amorphous support. However, image contrast alone is not sufficient to establish crystallinity in this system because local mass-thickness variations could also contribute to the darker intensity. To test more directly whether particle-free regions are  crystals, 4D-STEM is used to examine the 10~nm AlO$_x$ samples heated at 750 and 800$^\circ$C, and construct virtual crystallinity maps from those datasets (Fig.~S4).

The virtual crystallinity maps were generated using the ratio of the intensity of the second-brightest diffraction spot to that of the central beam in each diffraction pattern. Under this definition, amorphous regions give a value of zero because no diffraction spot other than the central beam is present. Although some crystalline regions may still appear weak in this metric if their local orientation lies far from a zone axis, nonzero crystallinity values provide direct evidence for the presence of crystalline material. As expected, strong signals of crystallinity are observed at the locations occupied by Pt particles in the post-heating state, as crystalline Pt itself produces intense diffraction spots in the 4D-STEM patterns. Importantly, however, both the 750 and 800$^\circ$C maps also show nonzero crystallinity in regions of the support where Pt particles are absent, confirming that the particle-free and crystallized support regions spatially coincide. This provides additional support for propagating particle-free regions in Fig.~\ref{fig3} as signatures of an advancing crystallization interface.

Following Eq.~\ref{eqn_veli} and the configurational force, a gradient in surface chemical potential induced by substrate heterogeneity is believed to drag particles during the transformation. The surface chemical potential gradient can be estimated by the difference in equilibrium contact angle between the different substrate regions (gradients in stress or other thermodynamics variables and gradients in the substrate composition may also contribute). The platinum nanoparticles maintain an isotropic shape during the transformation, therefore the platinum-vacuum surface energy is believed to be constant during the motion and the adhesion energy ($W^{ad}$) difference sets the angle difference, $\gamma^{v}_{NP} (\cos(\theta_{Y, +\hat{x}}) - \cos(\theta_{Y, -\hat{x}})) = W^{ad}_{+\hat{x}} - W^{ad}_{-\hat{x}}$. The adhesion energy between platinum and amorphous AlO$_x$ is difficult to measure due to the variability in amorphous structure. One might suspect a higher adhesion energy between platinum and the crystalline phase, due to the higher surface energy of the crystalline phase ($\gamma_{Al_2O_3} = 1.52 J/m^2$ \cite{castillo2003} and $\gamma_{AlO_x} > 0.97 J/m^2$ \cite{tavakoli2013amorphous}). However, owing to the favorable platinum-oxygen bonding \cite{Oware2021}, we suspect the platinum-crystalline interface has a higher interfacial energy compared to the platinum-AlO$_x$ interface (thus lower adhesion energy). Macroscopic adhesion energy measurements of blistering of Pt thin films on ALD deposited amorphous alumina corroborate our hypothesis of higher adhesion energy \cite{berdova2013}. Consequently, with AlO$_x$ expected to have a lower equilibrium contact angle compared to Al$_2$O$_3$, Eq.~\ref{eqn_veli} predicts motion towards the amorphous phase.

In Fig.~\ref{fig4}, we further distill the governing physics by using a diffuse interface (phase field) model capturing phase transformation kinetics, mass diffusion, and simultaneously evolving substrate-surface energy gradients \cite{chen2025interfacial}. A solid particle is placed on a region of a substrate representing amorphous AlO$_x$. Another region with larger equilibrium contact angle (set following our previous discussion by decreasing the substrate-particle adhesion energy) is associated with $\gamma$-Al$_2$O$_3$ support phase. By assuming a planar substrate with an interface between the amorphous and crystalline phase moving at a constant velocity, the boundary between $\gamma$-Al$_2$O$_3$ and AlO$_x$ is shifted a pixel every $10^5$ time steps. Fig.~\ref{fig4}(a-c) illustrate density map snapshots during the crystallization of the support, where the black vertical line is the interface between the amorphous and crystalline phases with the red arrow showing the direction of motion. Within the crystallization front, the particle moves towards the front, where the difference in contact angles gradually increases. As the front is underneath the particle, the surface energy gradients induce mass flow along the surface to drive particle migration toward the side of lower equilibrium contact angle. This process occurs due to curvature gradients along the particle surface; see Fig.~\ref{fig4}(d). The particle center of mass and the crystallization front location are presented in Fig.~\ref{fig4}(e), in which the particle center of mass is found to accelerate to a near steady-state velocity that is slightly lower than that of the crystallization front.

Our findings suggest a dynamic balance between the front velocity and the rate at which the particle can relax the energy state for a given front position. When the particle relaxation rate is fast relative to the front velocity, particle migration appears as stick-slip, pinned at the front location, moving toward the higher-affinity side, awaiting further front motion. Meanwhile, when the front velocity is faster than the density redistribution, the particle will be left behind, with the particle curvature adapting to the new substrate only at late times. Fig.~\ref{fig4} presents the intermediary case where a particle curvature gradually develops, effectively dragging the particle along with the front. The phenomenon is a function of particle size, since a larger particle radius increases the timescale for mass redistribution. 

To quantify the crystallization front propagation velocity, we tracked the migration of individual Pt particles during the early stage of front propagation on the 20~nm AlO$_x$ film and extracted migration velocities from linear fits to their cumulative displacement curves (Fig.~S5). The mean front velocity estimated from particle tracking is $1.28 \pm 0.48$~nm/s, where the spread across particles reflects spatial heterogeneity in the local front propagation speed as the front does not reach each particle simultaneously. If driven by curvature gradients, our earlier analysis (Eq.\ref{eqn_veli}) suggests a curvature gradient coefficient, $\Gamma = \mathcal{O}(10^{-5})$. The magnitude of $\Gamma$ points towards a small difference in curvature forming on the platinum nanoparticle straddling both the amorphous AlO$_x$ and $\gamma$-Al$_2$O$_3$ support. We note that differences may arise at the contact lines, or along the nanoparticle/substrate interface, which can contribute additional drag to particle motion. Furthermore, we have highlighted curvature gradients as being the main driver, but note that other gradients, such as mechanical stress imbalance, could also play a role.

In relation to our experimental findings, rapid nucleation of the crystalline phase may have an elevated front velocity compared to Pt mass redistribution. As a consequence, particles are left behind during the initial events. Once the front reaches near-steady state, particles are dragged along and can coalesce with other particles along the way. When particles reach large sizes from repeated coalescence events, we suspect the front velocity may again exceed the critical threshold and particles are left in the wake of the front. In our experiments, however, crystallization fronts impinge before the critical sizes are reached, resulting in particle distributions resembling the underlying substrate grain network. Thus, our results demonstrate the possibility of patterning catalytic particles by control over coupled kinetics.

In conclusion, our \textit{in situ} TEM observations, supported by 4D-STEM analysis and phase-field simulations, demonstrate that propagating crystallization fronts in amorphous AlO$_x$ actively drag supported Pt nanoparticles through contact-line asymmetry imposed by the advancing boundary. The driving force acts through the asymmetry in the interfacial energy of the particles and the substrate at the advancing and trailing contact lines, creating curvature gradients along the surface of the particles that redistribute mass and displace the center of mass of the particles. The competition between the crystallization front velocity and the timescale for particle mass redistribution through surface and bulk diffusion determines the outcome: particles are dragged with the front when these rates are comparable, and left behind when the front advances too rapidly. The alumina crystallization and Pt nanoparticle system studied here is one realization of a more general principle: any propagating surface-energy boundary on a substrate---whether generated by an amorphous-to-crystalline transition, a structural phase transformation, or grain boundary migration---can serve as a deterministic driver of supported nanoparticle transport.

\section{Acknowledgments}

This study was supported by the National Science Foundation, Division of Materials Research, Metals and Metallic Nanostructures program, under grant DMR-2303084. This work was performed in part at the University of Pennsylvania's Singh Center for Nanotechnology, an NSF National Nanotechnology Coordinated Infrastructure (NNCI) member supported under Grant NNCI-2025608 and ECCS-1542153, and through the use of facilities supported by the University of Pennsylvania Materials Research Science and Engineering Center (MRSEC) under grant DMR-2309043. We thank Lucy Decker, Shengshong Yang, and Christopher Murray of the University of Pennsylvania Department of Chemistry for assistance with nanoparticle synthesis and self-assembly.

\bibliography{ref_20260421}

\end{document}